\documentclass{article} %
\usepackage{colm2024_conference}

\usepackage{booktabs}
\usepackage{graphicx}
\usepackage{enumitem}
\usepackage[T1]{fontenc}
\usepackage{caption}
\usepackage{multirow}
\usepackage{tabularx}
\usepackage{makecell}
\usepackage{listings}
\usepackage{adjustbox}
\usepackage{placeins}
\usepackage{flafter}
\usepackage{xurl}

\usepackage{amsmath,amsfonts,bm}

\def\eqref#1{equation~\ref{#1}}

\def\1{\bm{1}}

\DeclareMathAlphabet{\mathsfit}{\encodingdefault}{\sfdefault}{m}{sl}
\SetMathAlphabet{\mathsfit}{bold}{\encodingdefault}{\sfdefault}{bx}{n}

\begin{document}
\begin{center}
\includegraphics[width=0.31\textwidth]{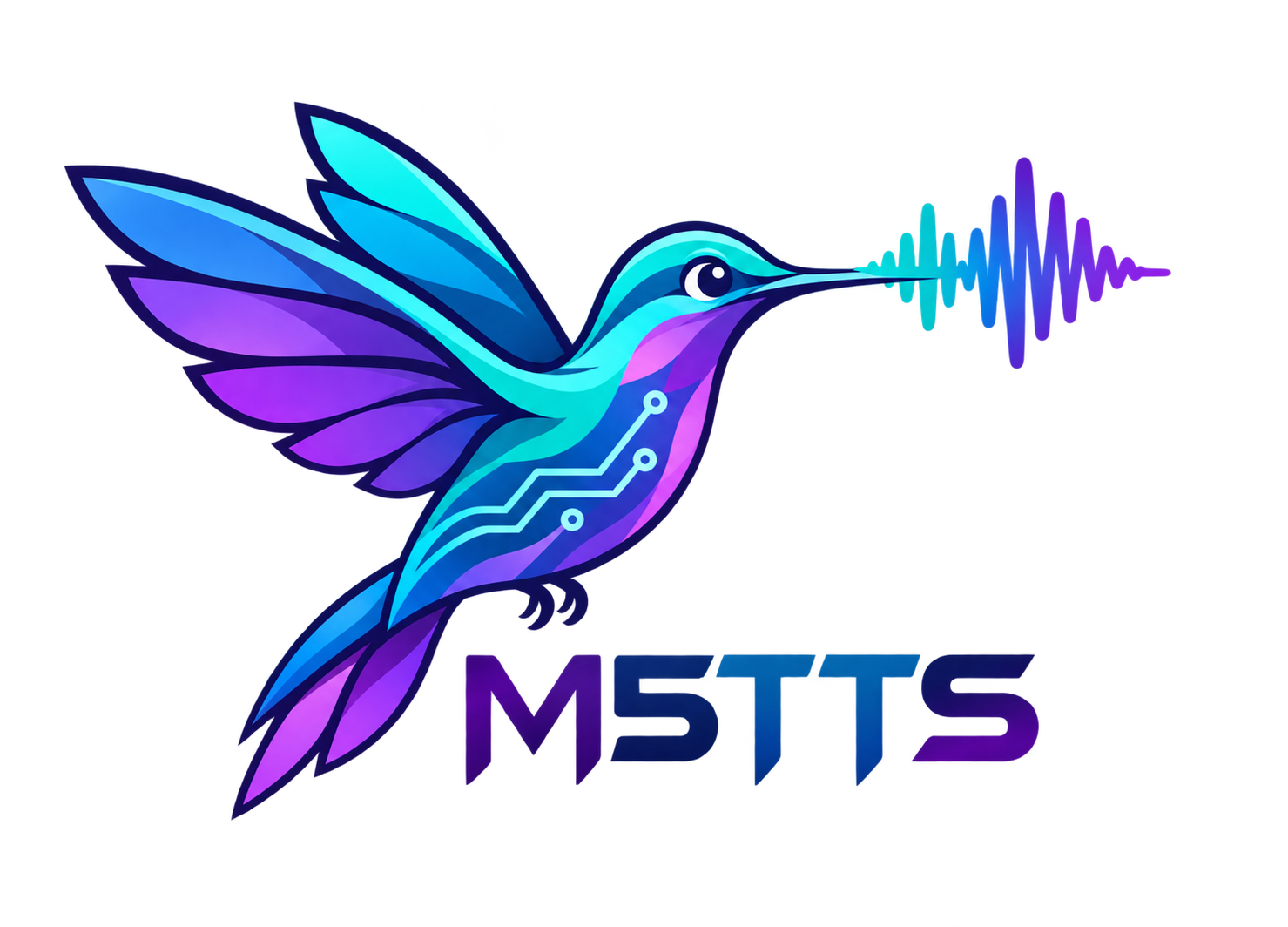}
\end{center}
\title{M5-TTS: A Mini-Scale Multilingual and Multi-Dialect Text-to-Speech Language Model with Multi-Token Prediction}

\author{
Wenhao Li$^{1,*}$, Jingbin Hu$^{1,*}$, Shengming Ling$^{1,*}$, Xiaming Ren$^{1,*}$, Haoyu Zhang$^{1,*}$,  \\ Jinrui Liang$^{1}$, Hanke Xie$^{1}$, Qirui Zhan$^{1}$, Guobin Ma$^{1}$, Huakang Chen$^{1}$, \\ Chunbo Hao$^{1}$, Jianguo Zheng$^{1}$, Wenkang Hou$^{1}$, Dake Guo$^{1}$, Yuepeng Jiang$^{1}$, \\ Linhan Ma$^{1}$, Yunxiang Chen$^{2}$, Houdun Liu$^{2}$, Su Feng$^{2}$, Chuang Ding$^{3}$, Pengcheng Zhu$^{4}$, Liumeng Xue$^{5,\dagger}$, Lei Xie$^{1,\dagger}$
\\[0.5em]
$^{1}$Audio, Speech and Language Processing Group (ASLP@NPU),
School of Computer Science, Northwestern Polytechnical University\\
$^{2}$Shenzhen Pimei Technology Co., Ltd., $^{3}$Moonstep AI, $^{4}$WeNet Open Source Community \\
$^{5}$Audio, Speech and Language Processing Group (ASLP@NJU),
School of Intelligence Science and Technology, Nanjing University
}

\newcommand{\tabincell}[2]{\begin{tabular}{@{}#1@{}}#2\end{tabular}}
\newcommand{\fix}{\marginpar{FIX}}
\newcommand{\new}{\marginpar{NEW}}
\newcommand{\jin}[1]{{\color{blue}#1}}

\maketitle
\begingroup
\renewcommand{\thefootnote}{\fnsymbol{footnote}}
\footnotetext[1]{These authors contributed equally to this work.}\footnotetext[2]{Corresponding authors.}
\endgroup

\begin{abstract}
We present \textbf{M5-TTS}, a compact text-to-speech system supporting
multilingual and Chinese multi-dialect synthesis, zero-shot voice cloning,
speaker-specific multilingual generation, and CPU streaming. Its
\textbf{50M-parameter language model}, excluding the speech codec, combines
a temporal backbone with a local multi-token prediction module over
16 codebooks. GRPO post-training normalizes speaker-similarity and pause
rewards separately within each group before weighting, then adds
CER-guided local process feedback derived from character-level errors,
with backward propagation within each clause and projection onto codec frames.
M5-TTS achieves 1.37\% CER on Seed-ZH and 2.91\% WER on Seed-EN,
corresponding to relative error reductions of 62.2\% and 35.3\% over
MOSS-TTS-Nano with half the language-model parameters. It also achieves
lower content error rates in five of six evaluated MiniMax languages,
while difficult Chinese inputs remain a limitation.
On four Intel Xeon Gold 6530 cores, it achieves a real-time factor of
0.653 and a warm first-chunk latency of 225\,ms.
VoiceJudge, our internal model for assessing speech naturalness and
expressiveness, prefers M5-TTS over MOSS-TTS-Nano in 84.89\% and
86.50\% of naturalness comparisons under the two A/B presentation orders.
We will release inference code, together with
voice-cloning and speaker-specific checkpoints, to support reproducible
research on compact speech generation for on-device voice interaction.
\end{abstract}

\section{Introduction}
\label{sec:intro}
Speech generation is a key component of spoken interaction for embodied agents, personal assistants, and intelligent devices. In these settings, a speech synthesizer needs to produce intelligible and natural speech, maintain a consistent voice, and begin playback with a short delay. Multilingual and dialectal capabilities further allow a single device to communicate with users across linguistic communities. Local inference can reduce dependence on network connectivity and give developers greater control over the speech generation pipeline. These requirements motivate compact text-to-speech (TTS) systems that balance synthesis quality, linguistic coverage, voice flexibility, and streaming efficiency under limited computational resources.

Neural TTS research has explored diverse generation architectures. Early end-to-end approaches, such as VITS~\cite{vits}, combine variational inference and adversarial learning for waveform generation, while non-autoregressive models such as FastSpeech 2~\cite{fastspeech2} improve synthesis efficiency through parallel acoustic modeling and explicit duration, pitch, and energy conditioning. However, point-wise acoustic regression objectives used in some non-autoregressive systems can encourage over-smoothing~\cite{ren2022revisiting}, reducing acoustic and prosodic variation. Richer generative models aim to capture this variation more effectively.

Language-model-based TTS approaches use speech tokens as intermediate
representations~\cite{valle}, enabling autoregressive speech generation
through sequence modeling. Subsequent studies have explored diverse speech representations and generation paradigms. CosyVoice 2~\cite{cosyvoice2} combines autoregressive language modeling with conditional flow matching, while Qwen3-TTS~\cite{qwentts} employs multi-codebook tokens for multilingual and streaming synthesis. Recent work has also explored continuous speech generation without relying on conventional discrete tokens. DiTAR~\cite{ditar} combines an autoregressive language model that captures dependencies across speech patches with a local diffusion Transformer that generates continuous acoustic features within each patch. VoxCPM2~\cite{voxcpm2} further explores tokenizer-free diffusion autoregressive modeling for multilingual speech generation and controllable voice cloning. Despite these advances, the substantial model sizes and inference costs of these systems can make low-latency streaming challenging on resource-constrained edge devices, particularly under CPU-only inference.

The growing demand for on-device voice interaction and embodied
intelligence therefore creates a pressing need for compact TTS models
that preserve naturalness and expressiveness while supporting
efficient, responsive streaming synthesis. Recent compact TTS systems have further advanced on-device speech generation. MOSS-TTS-Nano~\cite{mossttsnano} explores lightweight language-model-based speech generation with discrete codec tokens, while Pocket TTS~\cite{pockettts} demonstrates efficient CPU-based inference with compact continuous models. Other lightweight projects, including Audio8-TTS Preview~\cite{audio8tts_preview2026}, Kokoro-TTS~\cite{kokoro2025}, Hojo-TTS-Light~\cite{hojottslight}, TinyTTS~\cite{tinytts}, and sanoTTS~\cite{sanotts}, further investigate reducing model complexity and computational requirements for resource-constrained speech synthesis. 

Combining broad language and dialect coverage, zero-shot voice cloning,
and CPU streaming within a 50M-parameter language model remains a
challenging design objective. We study this combination under a shared,
compact generation architecture.

We present \textbf{M5-TTS}, a \textbf{M}ini-scale \textbf{M}ultilingual and \textbf{M}ulti-dialect TTS
language \textbf{M}odel with \textbf{M}ulti-token prediction (MTP), built around 50M parameters
excluding the speech codec. A temporal backbone and a local Transformer
jointly generate a 16-codebook speech representation. Language and accent
conditioning support linguistic control, while reference speech enables
zero-shot cloning and speaker-specific fine-tuning provides a dedicated
multilingual voice. To refine synthesis, we use Group Relative Policy
Optimization (GRPO)~\cite{shao2024deepseekmath} with two complementary
feedback paths. Global SIM and pause rewards assess overall speaker
resemblance and phrasing; each reward is normalized within its candidate
group before weighted aggregation. CER-guided local process feedback
compares character edit counts across candidate utterances, propagates
negative advantages backward within a clause, and projects them onto
codec frames. The resulting local process advantage is added to the global
advantage and shared by all codec tokens in each frame, providing content
feedback for the generation decisions associated with each text region.

Our experiments examine synthesis quality and streaming performance, with
MOSS-TTS-Nano~\cite{mossttsnano} as the main comparison among lightweight
language-model-based TTS systems. Using half as many language-model
parameters, M5-TTS achieves 1.37\% CER on Seed-ZH and 2.91\% WER on
Seed-EN, compared with 3.62\% and 4.50\% for MOSS-TTS-Nano, respectively.
On four cores of an Intel Xeon Gold 6530, M5-TTS achieves a real-time
factor of 0.653 and a warm first-chunk latency of 225\,ms.

To support repeated perceptual comparisons without collecting new human
ratings for every experiment, we introduce \textbf{VoiceJudge}, an internal
reward model combining self-supervised speech representations with a
Transformer-based scoring module. It is trained on approximately 100,000
human-annotated preference pairs and supports multidimensional A/B
assessment, including naturalness and expressiveness.
Here we use its naturalness dimension on a strictly deduplicated test
set that does not overlap with its training data.
VoiceJudge prefers M5-TTS over MOSS-TTS-Nano in 84.89\% and 86.50\%
of 311 valid comparisons under original and reversed A/B orders,
with 92.60\% agreement on the winning system across orders.

We will release inference code, together
with zero-shot voice-cloning and speaker-specific checkpoints, and plan
to release VoiceJudge model weights and inference code to support further
community development. With this work, we aim to provide a practical
speech generation component for on-device applications and embodied
voice interaction.

\section{Architecture}
\label{sec:architecture}

\subsection{Tokenizer}

We adopt Qwen-TTS-Tokenizer-12Hz from Qwen3-TTS~\cite{qwentts} to convert speech into discrete representations at 12.5 frames per second. Each frame contains 16 codebook indices, comprising one semantic codebook and 15 acoustic codebooks. The encoder consists of a downsampling convolutional network followed by a windowed Transformer. Its latent features are quantized through a semantic VQ branch and a 15-stage acoustic RVQ branch, which represents complementary acoustic information. A windowed Transformer and an upsampling convolutional network then reconstruct the waveform from the quantized representations.
\begin{figure}[!htbp]
    \centering
    \includegraphics[width=0.8\textwidth]{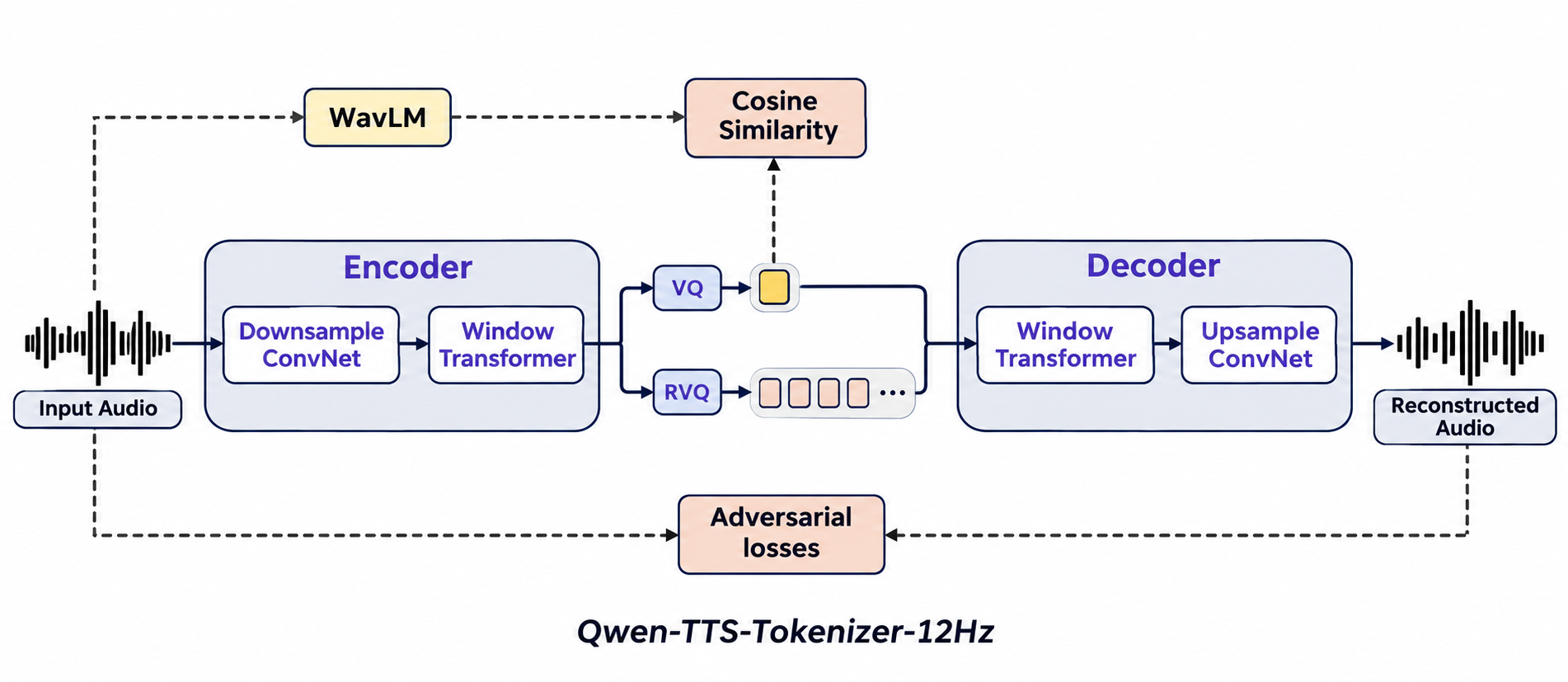}
    \caption{Qwen-TTS-Tokenizer-12Hz~\cite{qwentts} adopted by M5-TTS: a semantic VQ branch and an acoustic RVQ branch provide 16 codec tokens per frame for causal waveform reconstruction.}
    
    \label{fig:tokenizer}
\end{figure}

In the tokenizer's original training framework~\cite{qwentts}, WavLM~\cite{wavlm2021} provides teacher features for semantic distillation of the first codebook using a cosine-similarity objective. Adversarial training encourages perceptually faithful reconstruction, while a multi-scale mel-spectrogram loss constrains spectral consistency. Both the encoder and decoder operate causally: speech tokens are extracted without look-ahead, and the corresponding waveform is reconstructed incrementally. This combination of a low frame rate and causal reconstruction provides M5-TTS with a compact speech representation suited to streaming generation.

\FloatBarrier
\subsection{Backbone}

Following the hierarchical Transformer organization of Marvis-TTS~\cite{marvis2025} and the multi-codebook prediction scheme of Qwen3-TTS~\cite{qwentts}, M5-TTS uses a compact language-modeling architecture with 50M parameters, excluding the speech codec. The architecture comprises a causal backbone Transformer for temporal modeling and a lightweight local Transformer serving as the MTP module. The conditioning interface orders accent, speaker, and language IDs before the text sequence; the speaker ID is used in speaker-specific fine-tuning, whereas pre-training uses accent and language IDs. For each historical audio frame, codebook-specific embeddings of its 16 codec tokens are summed into a single frame-level input, allowing the backbone to model speech at the codec frame rate.

\begin{figure}[!htb]
    \centering
\includegraphics[width=0.8\textwidth]{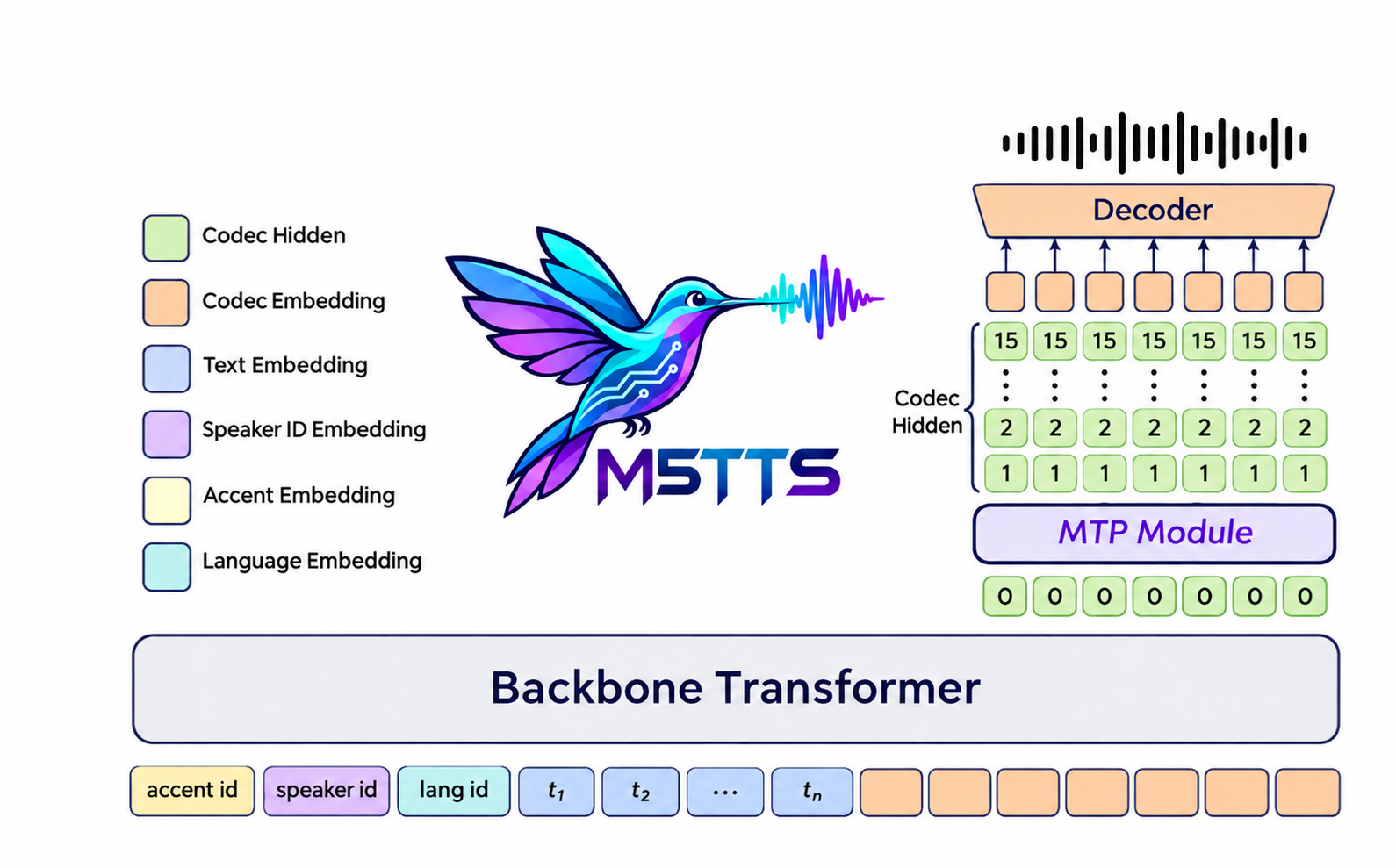}
    \caption{M5-TTS architecture. The temporal backbone predicts codebook 0; the local MTP module generates codebooks 1--15 within each frame, which are then decoded into speech.}
    \label{fig:backbone}
\end{figure}

Generation follows a hierarchical codebook prediction scheme. At each time step, the backbone summarizes the conditioning sequence and previously generated audio frames into a hidden state. A dedicated output head first predicts the zeroth codebook token. The backbone state and this token's embedding are projected to the MTP input space. The MTP module then autoregressively generates codebooks 1--15, with each prediction conditioned on the preceding codebooks within the same frame. The completed frame is aggregated and fed back to the backbone for the next temporal step.

During inference, the backbone retains its attention cache across audio frames, while the MTP cache is reset for each new frame. This factorization separates temporal modeling from within-frame acoustic refinement and avoids expanding all 16 codebooks into separate backbone time steps. Completed codec frames are passed to the causal waveform decoder for incremental reconstruction, supporting streaming synthesis.

\section{Methods}
\label{methods}
Our method consists of three components: pre-training, post-training,
and efficiency optimization. We first establish multilingual and
multidialect speech generation through basic pre-training, followed by
continued pre-training on curated data to strengthen acoustic quality,
expressiveness, and pronunciation robustness. Post-training combines
GRPO, which adds CER-guided local process feedback to global
speaker-similarity and pause feedback, with speaker-specific fine-tuning
for dedicated multilingual voices. Finally, efficiency optimization
improves inference execution, cache reuse, and streaming scheduling
to support low-latency speech synthesis on CPUs.

\subsection{Pre-training}
\subsubsection{Basic Pre-training}

M5-TTS is pre-trained on approximately 400,000 hours of speech.
The corpus combines the multilingual Emilia and Emilia-YODAS
collections~\cite{emilia,emilia_large2025}, approximately 160,000 hours
of in-house recordings, and WenetSpeech4TTS, WenetSpeech-Yue, and
WenetSpeech-Chuan~\cite{wenetspeech4tts,wenet-yue,wenet-chuan}.
Together, these sources cover Chinese, English, German, French,
Japanese, and Korean, as well as regional Chinese speech.

Pre-training conditions on the accent ID, language ID, and target text, in that order. Speaker IDs are introduced during speaker-specific fine-tuning. Speech is represented by 16 codebook tokens per codec frame. Supervised training learns text--speech correspondences under the hierarchical autoregressive factorization: the temporal backbone predicts the zeroth codebook, and the local MTP module predicts the remaining codebooks conditioned on the backbone state and preceding within-frame tokens. Teacher forcing supplies the observed speech history during likelihood training. This establishes a general speech-generation policy for subsequent refinement of pronunciation, speaker consistency, and speaker-specific synthesis.

\subsubsection{Continued Pre-training}

To improve acoustic quality, expressiveness, and pronunciation
robustness in multilingual and multidialectal synthesis, we perform
continued pre-training (CPT) on a curated subset of the pre-training
corpus. This subset comprises approximately 65,000 hours of speech,
with selection prioritizing high-quality recordings, expressive speech,
and challenging samples involving complex pronunciation patterns.
The corpus covers Chinese, English, German, French, Japanese, and
Korean, together with dialect data including Cantonese and Sichuanese.
Language and dialect tags are encoded alongside the target text to
provide explicit conditioning for speech generation.

\subsection{Post-training}

\subsubsection{Group Relative Policy Optimization}
\label{sec:grpo}

\begin{figure}[!htb]
    \centering
    \includegraphics[width=\textwidth]{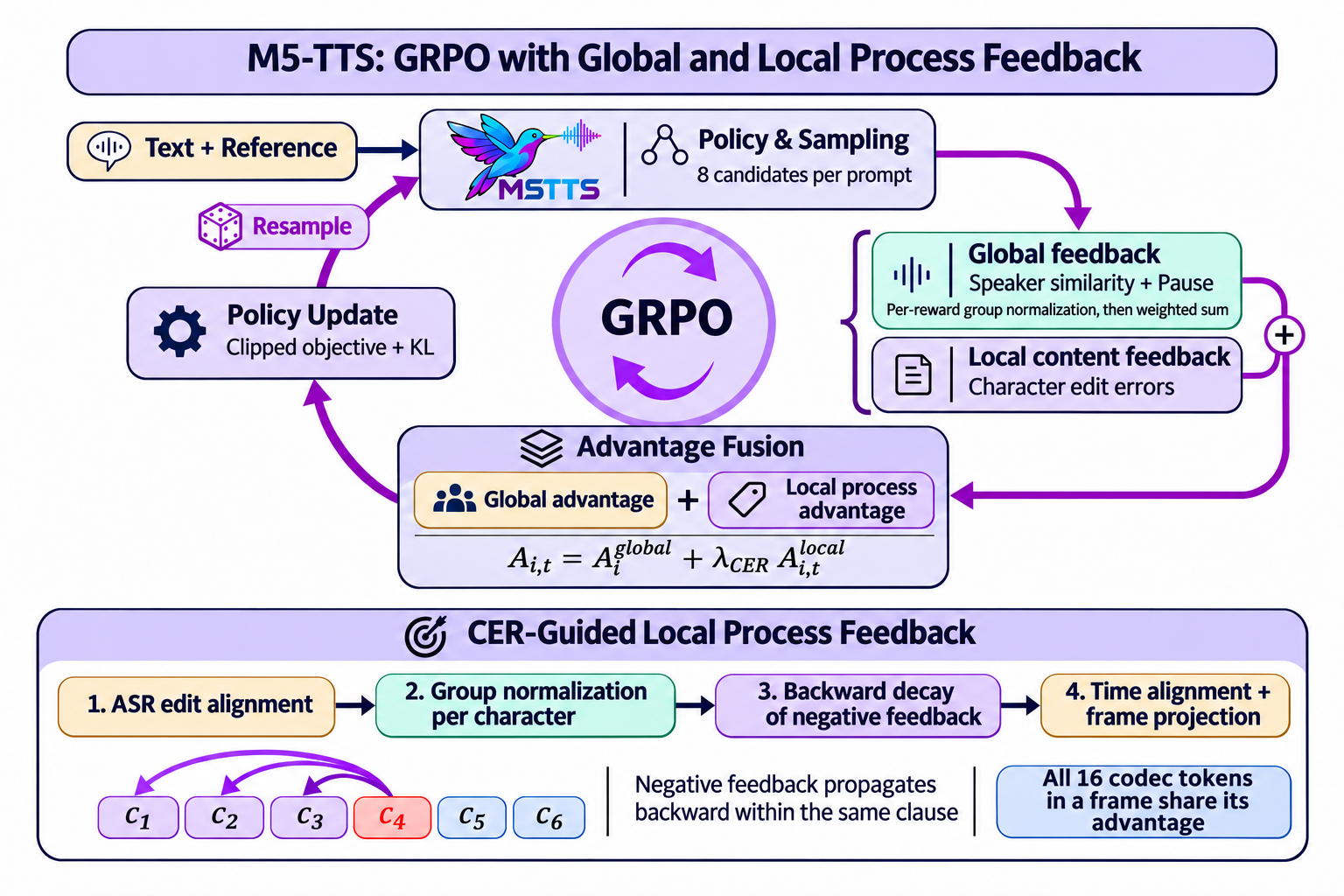}
    \caption{GRPO with global feedback and CER-guided local process feedback. Utterance-level
    speaker-similarity and pause rewards are normalized separately within
    each candidate group, then weighted and summed to obtain a shared
    global advantage.
    ASR edit alignment yields per-character group-relative advantages;
    negative advantages propagate backward within a clause before temporal
    projection onto codec frames. The local process advantage is added
    to the global advantage, and all 16 codec tokens in a frame share the
    resulting advantage for the policy update.}
    \label{fig:grpo}
\end{figure}

We use Group Relative Policy Optimization (GRPO)~\cite{shao2024deepseekmath}
to combine overall synthesis feedback with local feedback on content errors.
For the same target text and reference speech, the rollout policy samples
$G$ candidate utterances. Each candidate receives an utterance-level score
for speaker similarity and pausing, together with character-level feedback
derived from ASR errors. The global advantage is shared across the utterance;
CER-guided local process feedback adds a frame-specific advantage to it.
All codec tokens within the same frame share the resulting advantage.
Figure~\ref{fig:grpo} summarizes this two-branch procedure.

\paragraph{Global feedback.}
Speaker similarity measures resemblance to the reference voice. The pause
reward measures whether aligned inter-character gaps satisfy the duration
constraints for ordinary and punctuation-associated boundaries. We obtain
timestamps with Qwen3-ForcedAligner~\cite{qwen3asr2026}; the pause-score
definition is given in Appendix~\ref{app:grpo_details}.
We normalize each reward separately within the candidate group before
combining the resulting component advantages:
\begin{equation}
\begin{aligned}
    A_i^{k}
    &=\frac{R_i^{k}-\mu_k}{\sigma_k+\varepsilon},
      \qquad k\in\{\mathrm{SIM},\mathrm{pause}\},\\
    A_i^{\mathrm{global}}
    &=\lambda_{\mathrm{SIM}}A_i^{\mathrm{SIM}}
      +\lambda_{\mathrm{pause}}A_i^{\mathrm{pause}},
\end{aligned}
\label{eq:grpo-global-advantage}
\end{equation}
where $R_i^{\mathrm{SIM}}=\mathrm{SIM}_i$, and $\mu_k$ and $\sigma_k$
are the mean and standard deviation of reward $k$ over the $G$ candidates
for the same prompt. The constant $\varepsilon$ ensures numerical stability.
We use $\lambda_{\mathrm{SIM}}=1.0$ and
$\lambda_{\mathrm{pause}}=0.5$.
The weighted sum is used directly as the global advantage, without
further normalization. A reward that is identical across the group
contributes zero advantage.
Sentence-level CER is not broadcast as an additional global reward;
content feedback enters through the local branch below.

\paragraph{CER-guided local process feedback.}
We align each candidate's ASR transcript with the target text using edit
alignment. Let $e_{i,j}$ be the number of edit operations assigned to target
character $j$ in candidate $i$. Substitutions and deletions are assigned to
the affected character; insertions are assigned to the following target
character, or to the last character for trailing insertions.
These are character edit counts, not CER values computed for individual
codec tokens.
At each target position, we compare the edit counts across the $G$ candidates:
\begin{equation}
    a_{i,j}
    =\frac{\mu_j-e_{i,j}}{\sigma_j+\varepsilon},
\label{eq:grpo-char-advantage}
\end{equation}
where $\mu_j$ and $\sigma_j$ are the group mean and standard deviation
of the edit counts at position $j$.
Fewer errors than the group mean produce positive local feedback;
more errors produce negative feedback. If all candidates have the same
edit count at a position, its unpropagated local advantage is zero.

An error may depend on generation decisions made before its audible
location. We therefore propagate negative character advantages backward
within the same clause, with exponential decay over character distance:
\begin{equation}
    \widetilde{a}_{i,j}
    =a_{i,j}
     +\sum_{k\in\mathcal{C}_j^{+}}
       \exp\!\bigl(-(k-j)\bigr)\min(a_{i,k},0),
\label{eq:grpo-char-propagation}
\end{equation}
where $\mathcal{C}_j^{+}$ contains the characters after $j$ in its clause.
Only negative advantages are propagated. Earlier characters can thus
receive attenuated feedback from a later error, even when they were
recognized correctly.

\paragraph{Projection onto codec frames.}
Forced alignment provides timestamps for the recognized text, and edit
alignment associates these intervals with target characters. We project
the propagated advantages onto codec frames. For a character with an
acoustic interval, its projection weight $w_{i,t,j}$ is the overlap
duration with frame $t$, divided by the frame duration. If the weights
within a frame sum to more than one, they are normalized by that sum.
A frame entirely inside one character interval therefore receives the
full character advantage, rather than a share divided by the number of
frames in that interval. Deletions have no spoken interval, so their
feedback is attached to an available following boundary decision or
the terminal decision. The projection weights also encode these
deletion-to-boundary assignments.

The resulting local process advantage is added to the global advantage:
\begin{equation}
\begin{aligned}
    A_{i,t}^{\mathrm{local}}
       &=\sum_j w_{i,t,j}\,\widetilde{a}_{i,j},\\
    A_{i,t}
       &=A_i^{\mathrm{global}}
         +\lambda_{\mathrm{CER}}A_{i,t}^{\mathrm{local}},
         \qquad \lambda_{\mathrm{CER}}=0.5.
\end{aligned}
\label{eq:grpo-frame-advantage}
\end{equation}
All 16 codec tokens in a frame share $A_{i,t}$. Thus, every generated decision
receives the global synthesis feedback, while the local branch adds
position-dependent content feedback.

\paragraph{Policy optimization.}
For codebook token $o_{i,t,c}$ with conditioning context $h_{i,t,c}$,
the importance ratio compares the current policy with the rollout policy
that generated the candidate:
\begin{equation}
    r_{i,t,c}(\theta)
    =\frac{\pi_\theta(o_{i,t,c}\mid h_{i,t,c})}
           {\pi_{\mathrm{old}}(o_{i,t,c}\mid h_{i,t,c})}.
\label{eq:grpo-probability-ratio}
\end{equation}
We minimize a PPO-style clipped policy loss~\cite{schulman2017ppo}
with KL regularization:
\begin{equation}
\begin{aligned}
    \ell_{i,t,c}(\theta)
    &=-\min\!\left\{
       r_{i,t,c}(\theta)A_{i,t},\,
       \operatorname{clip}\!\left(r_{i,t,c}(\theta),
       1-\epsilon_-,1+\epsilon_+\right)A_{i,t}
       \right\},\\
    \mathcal{L}_{\mathrm{GRPO}}(\theta)
    &=\mathcal{L}_{\mathrm{clip}}(\theta)
      +\beta\mathcal{L}_{\mathrm{KL}}(\theta),
      \qquad \beta=0.01.
\end{aligned}
\label{eq:grpo-loss}
\end{equation}
Here $\epsilon_-$ and $\epsilon_+$ are the clipping margins, and
$\mathcal{L}_{\mathrm{clip}}$ aggregates the token losses $\ell_{i,t,c}$.
Ratios and clipping are computed separately for each codebook.
The KL term constrains deviation from a frozen supervised reference
policy $\pi_{\mathrm{ref}}$, which is distinct from $\pi_{\mathrm{old}}$.
Advantages and reference predictions are detached during optimization.
Loss aggregation and the rollout configuration are described in
Appendices~\ref{app:grpo_details} and~\ref{app:grpo_hyperparameters}.

\subsubsection{Speaker-Specific Fine-Tuning}

Speaker-specific fine-tuning adapts the shared model to curated recordings
of a designated speaker. A speaker identifier is added between the accent
and language identifiers, retaining the hierarchical token-prediction
objective. Curated recordings from the M2VoC Challenge
2021~\cite{m2voc} provide speaker-specific supervision.

This setting evaluates an adapted voice rather than matching a supplied
reference speaker at inference time. We therefore report custom-voice
and zero-shot voice-cloning results separately.

\subsection{Efficiency Optimization}
To enable low-latency streaming speech synthesis on both CPUs and GPUs, we optimize the inference pipeline of M5-TTS without modifying its hierarchical autoregressive architecture. Our optimizations focus on four aspects: computation graph optimization, stateful inference and cache reuse, adaptive streaming scheduling, and hardware-aware runtime optimization.

\subsubsection{Computation Graph Optimization}

We reduce inference overhead by restructuring the execution graphs of the backbone and depth decoder. After the backbone predicts the first codebook, we encapsulate prediction and sampling of the remaining 15 codebooks within a single local-decoder frame graph, reducing Python-level iteration and runtime invocation overhead. Autoregressive dependencies among these codebooks are preserved.

We further fuse the query, key, and value projections in attention layers, as well as the gate and up projections in feed-forward layers, to reduce redundant operations and intermediate memory accesses. In addition, we precompute the linear projections of audio embeddings, replacing repeated embedding projection operations with efficient lookup-table accesses.

\subsubsection{Stateful Inference and Cache Reuse}

To eliminate redundant computation during autoregressive generation, we introduce explicit KV-cache management for both the backbone and depth decoder. On CPUs, these states are explicitly passed between ONNX Runtime~\cite{onnxruntime2018} inference calls, enabling incremental decoding and independent ONNX deployment.

We also implement a two-level reference caching mechanism. Reference audio is first encoded into reusable codec tokens, while the CPU inference pipeline additionally caches decoder states associated with complete reference inputs. A bounded LRU policy controls cache memory usage.

For streaming waveform generation, we implement a stateful codec decoder that maintains causal convolution histories, transposed-convolution overlap states, and sliding-window attention KV caches. This allows subsequent decoding steps to process newly generated codec tokens while reusing historical states, reducing redundant computation across overlapping audio chunks.

\subsubsection{Adaptive Streaming Scheduling}

We adopt a latency-aware streaming strategy that separates initial-chunk generation from subsequent decoding. A smaller initial chunk is used to reduce time to first audio, while larger decoding batches improve throughput during continuous generation.

For CPU inference, the decoding batch size is dynamically selected from 4, 8, and 16 frames according to the amount of generated audio available ahead of playback. This adaptive scheduling strategy balances first-chunk latency, overall real-time factor, and playback continuity.

\subsubsection{Hardware-Aware Runtime Optimization}

We apply hardware-specific execution optimizations to improve inference efficiency across CPU and GPU platforms.

For CPU inference, we optimize ONNX Runtime execution through physical-core affinity, intra-operation thread tuning, single-threaded inter-operation scheduling, sequential execution, and graph optimization. We additionally reduce resource overhead through idle thread-pool suspension, memory-mapped embeddings, and preallocated generation buffers. The execution graph is selected according to the CPU and threading configuration.

For GPU inference, the TTS model operates in BF16 precision, while the codec decoder remains in FP32 to preserve numerical stability. We compile the depth-frame computation graph and codec decoding operations using \texttt{torch.compile}~\cite{ansel2024pytorch2}, with separate compiled codec graphs for fixed input-window shapes.

Finally, we reduce CPU--GPU synchronization and memory allocation overhead by accumulating EOS indicators within each decoding chunk and checking them at chunk boundaries, while preallocating codec-token, single-frame input, and positional buffers. The fast inference configuration additionally employs scaled dot-product attention (SDPA) for efficient attention computation.

\section{Experiments}
\label{sec:experiments}

\subsection{Benchmarks and Metrics}
Table~\ref{tab:benchmark_structure} summarizes the evaluation benchmarks.
We evaluate on Seed-TTS Eval~\cite{seedttseval2024}, the MiniMax TTS-Multilingual-Test-Set~\cite{minimaxspeech2025}, and the Cantonese and Sichuanese benchmarks from WenetSpeech-Yue and WenetSpeech-Chuan~\cite{wenet-yue,wenet-chuan}.

Content accuracy is measured by character error rate (CER), word error
rate (WER), or mixed error rate (MER), according to the benchmark.
Each error rate is $100(S+D+I)/N$, where $S$, $D$, and $I$ denote
substitutions, deletions, and insertions, and $N$ is the number of reference
units. We use CER for Mandarin, Japanese, Korean, and Sichuanese, and WER
for English, German, and French. For Cantonese, we report MER, treating
Chinese characters and English words as evaluation units, following
WenetSpeech-Yue~\cite{wenet-yue}. All error rates are reported in percent.
Speaker similarity (SIM) measures consistency with the reference voice,
and higher values are preferred.

For Seed-TTS Eval, we follow the official evaluation
scripts~\cite{seedttseval2024}, using Paraformer-zh~\cite{paraformer2022}
for Mandarin transcription and
Whisper-large-v3~\cite{radford2022robustspeechrecognitionlargescale}
for English transcription. Seed-TTS SIM is the cosine similarity between
embeddings extracted by WavLM-large fine-tuned for speaker
verification~\cite{wavlm2021}.

For MiniMax, Mandarin speech is transcribed with a locally loaded
SeACo-Paraformer~\cite{seaco_paraformer2023}, without auxiliary voice
activity detection (VAD) or punctuation-restoration models.
For English, German, French, Japanese, and Korean, we use OpenAI
Whisper-large-v3 with the target language explicitly specified.
SIM uses the same WavLM-large model and cosine-similarity procedure
as in Seed-TTS Eval.

\paragraph{Systems and settings.}
 We distinguish Voice Cloning and Custom Voice configurations throughout. The latter uses a specific speaker configuration and is not a matched zero-shot cloning comparison. Additional lightweight baselines are included where measurements are available. Missing entries indicate unavailable or unreported measurements; they do not establish that a capability is unsupported. Bold model names identify M5-TTS.

\providecommand{\mFiveMissing}{\textemdash}

\begin{table*}[!htbp]
\centering
\caption{Structure of the evaluation benchmarks.
CER, WER, and MER are reported in percent.}
\label{tab:benchmark_structure}

\small
\setlength{\tabcolsep}{7pt}
\renewcommand{\arraystretch}{1.15}

\begin{adjustbox}{max width=\textwidth}
\begin{tabular}{@{}lllll@{}}
\toprule
Benchmark & Subset & Language(s)
& Content metric $\downarrow$ & Speaker metric $\uparrow$ \\
\midrule

\multirow{3}{*}{Seed}
& Seed-ZH   & ZH & CER (\%) & SIM \\
& Seed-EN   & EN & WER (\%) & SIM \\
& Seed-Hard & ZH & CER (\%) & SIM \\

\midrule

\multirow{2}{*}{MiniMax}
& Multilingual & ZH, JA, KO & CER (\%) & SIM \\
& Multilingual & EN, DE, FR & WER (\%) & SIM \\

\midrule

\multirow{2}{*}{WenetSpeech}
& Chuan & Sichuanese & CER (\%) & SIM \\
& Yue &  Cantonese  & MER (\%) & SIM \\

\bottomrule
\end{tabular}
\end{adjustbox}
\par
\emph{Notes.}
ZH: Chinese; EN: English; DE: German; FR: French;
JA: Japanese; KO: Korean.

\end{table*}

\FloatBarrier
\subsection{Objective Results}

\paragraph{Seed-TTS evaluation.}
Table~\ref{tab:seed_results} shows that M5-TTS (GRPO) achieves
1.37\% CER on Seed-ZH and 2.91\% WER on Seed-EN, compared with
3.62\% and 4.50\% for MOSS-TTS-Nano. These correspond to relative
error reductions of 62.2\% and 35.3\%, respectively, with half the
language-model parameters. SIM is also higher on these subsets, at 0.65/0.53 versus 0.64/0.50. MOSS-TTS-Nano retains an advantage on Seed-Hard, with lower CER (15.77\% vs.\ 17.48\%) and higher SIM (0.63 vs.\ 0.59). Compared with the 170M-parameter Audio8-TTS Preview baseline, M5-TTS achieves lower Seed-Hard CER (17.48\% vs.\ 19.63\%), but has higher error rates on Seed-ZH and Seed-EN and lower SIM on all three subsets.

\begin{table*}[tbp]
\centering
\caption{Results on Seed-TTS Eval.
CER and WER are reported in percent.
Lower CER/WER and higher SIM indicate better performance.}
\label{tab:seed_results}

\small
\setlength{\tabcolsep}{5pt}
\renewcommand{\arraystretch}{1.17}

\begin{adjustbox}{max width=\textwidth}
\begin{tabular}{@{}lclrrrrrr@{}}

\toprule

\multirow{2}{*}{Model}
& \multirow{2}{*}{\makecell{Main Model \\ Parameters (M)}}
& \multirow{2}{*}{Setting}
& \multicolumn{2}{c}{Seed-ZH}
& \multicolumn{2}{c}{Seed-EN}
& \multicolumn{2}{c}{Seed-Hard} \\

\cmidrule(lr){4-5}
\cmidrule(lr){6-7}
\cmidrule(lr){8-9}

& & &
CER$\downarrow$ & SIM$\uparrow$
& WER$\downarrow$ & SIM$\uparrow$
& CER$\downarrow$ & SIM$\uparrow$ \\

\midrule

\textbf{M5-TTS (CPT)}
& 50 & Voice Cloning
& 1.70 & 0.64
& 3.60 & \underline{0.53}
& 20.92 & 0.57 \\

\textbf{M5-TTS (GRPO)}
& 50 & Voice Cloning
& 1.37 & \underline{0.65}
& 2.91 & \underline{0.53}
& 17.48 & 0.59 \\

\textbf{M5-TTS}
& 50 & Custom Voice
& \underline{1.26} & \mFiveMissing
& \underline{1.88} & \mFiveMissing
& 18.19 & \mFiveMissing \\

\midrule

Audio8-TTS Preview
& 170 & Voice Cloning
& \textbf{1.21} & \textbf{0.68}
& 1.96 & \textbf{0.57}
& 19.63 & \textbf{0.63} \\

MOSS-TTS-Nano
& 100 & Voice Cloning
& 3.62 & 0.64
& 4.50 & 0.50
& 15.77 & \textbf{0.63} \\

Pocket TTS
& 100 & Voice Cloning
& \mFiveMissing & \mFiveMissing
& \textbf{1.86} & 0.47
& \mFiveMissing & \mFiveMissing \\

Hojo-TTS-Light
& 80 & Voice Cloning
& 2.15 & 0.40
& 3.63 & 0.17
& 21.10 & 0.36 \\

Pocket TTS
& 100 & Custom Voice\textsuperscript{a}
& \mFiveMissing & \mFiveMissing
& 1.90 & \mFiveMissing
& \mFiveMissing & \mFiveMissing \\

Kokoro-TTS
& 82 & Custom Voice\textsuperscript{b}
& 1.40 & \mFiveMissing
& 2.04 & \mFiveMissing
& \textbf{6.33} & \mFiveMissing \\

Hojo-TTS-Light
& 40 & Custom Voice\textsuperscript{c}
& 1.78 & \mFiveMissing
& 2.97 & \mFiveMissing
& 22.70 & \mFiveMissing \\

Sano TTS
& 2.27 & Custom Voice\textsuperscript{d}
& 2.42 & \mFiveMissing
& 2.40 & \mFiveMissing
& \underline{7.08} & \mFiveMissing \\

Tiny TTS
& 1.6 & Custom Voice
& \mFiveMissing & \mFiveMissing
& 3.30 & \mFiveMissing
& \mFiveMissing & \mFiveMissing \\

\bottomrule
\end{tabular}
\end{adjustbox}

\par\vspace{4pt}
\begin{minipage}{\textwidth}
\footnotesize\raggedright
\emph{Notes.}
\mFiveMissing{} denotes an unavailable result or an unspecified setting.
Setting distinguishes Voice-Cloning and Custom-Voice configurations.
For models with multiple preset voices, Custom-Voice results are obtained
using one selected voice per language, as specified below.
Unmarked Custom-Voice models use a single fixed voice.
The Audio8-TTS row uses the Preview-0.1B
checkpoint~\cite{audio8tts_preview2026}.
Parameter counts are reported in millions and exclude speech codecs.
\par
\textsuperscript{a}
The English voice is recorded as \texttt{alba};
this spelling is preserved from the evaluation record.
\par
\textsuperscript{b}
The Chinese and English voices are recorded as \texttt{zf\_xiaobei}
and \texttt{af\_heart}, respectively.
\par
\textsuperscript{c}
The Chinese and English voices are recorded as \texttt{hojo\_zh\_f\_01}
and \texttt{hojo\_en\_m\_02}, respectively.
\par
\textsuperscript{d}
The Chinese and English voices are recorded as \texttt{Xiaoya}
and \texttt{Amy large}, respectively.
\end{minipage}
\end{table*}

\paragraph{Multilingual voice cloning.}
On MiniMax (Table~\ref{tab:minimax_results}), M5-TTS (GRPO) achieves
lower content error rates than MOSS-TTS-Nano in five of six languages.
The relative reductions are 42.7\%, 65.9\%, 56.3\%, 75.9\%, and 64.9\%
for English, German, French, Japanese, and Korean, respectively.
The largest reduction is observed in Japanese, where CER decreases
from 40.22\% to 9.70\%. Chinese remains a gap, with 2.05\% CER
versus 1.80\% for MOSS-TTS-Nano.
SIM is higher in German, French, and Japanese, but lower in Chinese,
English, and Korean.
Pocket TTS achieves lower error rates in its reported English,
German, and French voice-cloning evaluations, while M5-TTS has
higher SIM in English and German and matches French SIM at the
reported precision.

\begin{table*}[tbp]
\centering
\caption{Multilingual results on MiniMax.
CER and WER are reported in percent.
Lower CER/WER and higher SIM indicate better performance.}
\label{tab:minimax_results}

\small
\setlength{\tabcolsep}{4pt}
\renewcommand{\arraystretch}{1.17}

\begin{adjustbox}{max width=\textwidth}
\begin{tabular}{@{}ll*{6}{rr}@{}}
\toprule

\multirow{2}{*}{Model}
& \multirow{2}{*}{Setting}
& \multicolumn{2}{c}{ZH}
& \multicolumn{2}{c}{EN}
& \multicolumn{2}{c}{DE}
& \multicolumn{2}{c}{FR}
& \multicolumn{2}{c}{JA}
& \multicolumn{2}{c}{KO} \\

\cmidrule(lr){3-4}
\cmidrule(lr){5-6}
\cmidrule(lr){7-8}
\cmidrule(lr){9-10}
\cmidrule(lr){11-12}
\cmidrule(lr){13-14}

& &
CER$\downarrow$ & SIM$\uparrow$
& WER$\downarrow$ & SIM$\uparrow$
& WER$\downarrow$ & SIM$\uparrow$
& WER$\downarrow$ & SIM$\uparrow$
& CER$\downarrow$ & SIM$\uparrow$
& CER$\downarrow$ & SIM$\uparrow$ \\

\midrule

\textbf{M5-TTS (CPT)}
& Voice Cloning
& 2.54  & 0.70
& 3.67  & 0.62
& 4.60  & \textbf{0.63}
& 8.15  & 0.51
& 12.71 & \textbf{0.72}
& 3.10  & 0.66 \\

\textbf{M5-TTS (GRPO)}
& Voice Cloning
& 2.05  & 0.71
& 3.06  & 0.63
& 3.81  & \textbf{0.63}
& 8.14  & \textbf{0.53}
& 9.70  & \textbf{0.72}
& 2.88  & 0.66 \\

\textbf{M5-TTS}
& Custom Voice
& 1.16  & ---
& 2.24  & ---
& \textbf{0.70} & ---
& 5.33  & ---
& 10.27 & ---
& \textbf{2.09} & --- \\

\midrule

MOSS-TTS-Nano
& Voice Cloning
& 1.80  & \textbf{0.73}
& 5.34  & \textbf{0.65}
& 11.18 & 0.54
& 18.64 & 0.51
& 40.22 & 0.70
& 8.21  & \textbf{0.67} \\

Pocket TTS
& Voice Cloning
& \mFiveMissing & \mFiveMissing
& 2.41 & 0.61
& 3.50 & 0.54
& 6.71 & \textbf{0.53}
& --- & ---
& --- & --- \\

Pocket TTS
& Custom Voice\textsuperscript{a}
& ---  & ---
& 1.99 & ---
& 4.14 & ---
& 7.88 & ---
& ---  & ---
& ---  & --- \\

Sano TTS
& Custom Voice\textsuperscript{b}
& 1.41 & ---
& \textbf{1.85} & ---
& 3.18 & ---
& 8.91 & ---
& ---  & ---
& ---  & --- \\

Kokoro-TTS
& Custom Voice\textsuperscript{c}
& \textbf{1.03} & ---
& 2.58 & ---
& ---  & ---
& \textbf{4.79} & ---
& \textbf{3.72} & ---
& ---  & --- \\

\bottomrule
\end{tabular}
\end{adjustbox}

\par\vspace{3pt}
\begin{minipage}{\textwidth}
\footnotesize\raggedright
\textit{Notes.} --- denotes an unavailable result or an unspecified setting.
ZH, JA, and KO use CER; EN, DE, and FR use WER.
All values are rounded to two decimal places.
Setting distinguishes Voice-Cloning and Custom-Voice configurations.
For models with multiple preset voices, Custom-Voice results are obtained
using one selected voice per language, as specified below.
Unmarked Custom-Voice models use a single fixed voice.
\par
\textsuperscript{a}
The English, German, and French voices are recorded as
\texttt{alba}, \texttt{juergen}, and \texttt{estelle}, respectively.
\par
\textsuperscript{b}
The Chinese, English, German, and French voices are recorded as
\texttt{Xiaoya}, \texttt{Amy large}, \texttt{Thorsten}, and \texttt{Siwis},
respectively.
\par
\textsuperscript{c}
The Chinese, English, French, and Japanese voices are recorded as
\texttt{zf\_xiaobei}, \texttt{af\_heart}, \texttt{ff\_siwis}, and
\texttt{jf\_alpha}, respectively.
\end{minipage}

\end{table*}

\paragraph{Effect of GRPO post-training.}
Relative to CPT, GRPO reduces error rates on all three Seed subsets:
CER/WER decreases from 1.70\% to 1.37\% on Seed-ZH, from 3.60\%
to 2.91\% on Seed-EN, and from 20.92\% to 17.48\% on Seed-Hard.
The corresponding relative reductions are 19.4\%, 19.2\%, and 16.4\%.
On MiniMax, error rates decrease in all six languages, with the
largest relative reduction in Japanese (23.7\%); French changes
only marginally, from 8.15\% to 8.14\% WER.
Across both benchmarks, SIM is maintained or improved at the
displayed precision. These comparisons evaluate the complete post-training
recipe; they do not isolate the effects of the pause reward or
CER-guided local process feedback.

\paragraph{Speaker-specific synthesis.}
The custom-voice variant achieves 1.26\% CER on Seed-ZH and 1.88\%
WER on Seed-EN. Compared with the GRPO voice-cloning variant, it also has lower content error rates in five of six MiniMax languages, including 0.70\% WER in German. The GRPO voice-cloning variant performs better on Seed-Hard and MiniMax Japanese. Custom-voice results reflect dedicated speaker configurations and are reported separately from zero-shot voice cloning.

\paragraph{Dialect synthesis.}
Table~\ref{tab:dialect_results} evaluates Cantonese and Sichuanese,
using the same ASR evaluator for all systems within each dialect.
We use U2pp-Conformer-Yue~\cite{wenet-yue} for Cantonese and
FireRedASR-AED~\cite{xu2025fireredasropensourceindustrialgrademandarin}
for Sichuanese.
For Cantonese, M5-TTS achieves the lowest MER among the compared
systems on both subsets, at 8.55\% on Easy and 8.70\% on Hard.
Relative to CosyVoice2-Yue~\cite{cosyvoice2,wenet-yue}, these correspond to reductions of
17.2\% and 8.3\%; relative to Llasa-1B-Yue~\cite{ye2025llasa,wenet-yue}, the reductions are
21.5\% and 31.9\%.
Its SIM is lower than both baselines on both subsets, indicating
a remaining gap in speaker preservation.
For Sichuanese, M5-TTS reduces Hard CER from 8.78\% to 7.78\%
relative to CosyVoice2-WSC~\cite{cosyvoice2,wenet-chuan}, an 11.4\% relative reduction, but has
higher Easy CER (5.08\% vs.\ 4.28\%) and lower SIM on both subsets.
These results demonstrate strengths in content accuracy on
Cantonese and the Sichuanese Hard subset, alongside remaining
limitations in speaker similarity.

\begin{table*}[!htbp]
\centering
\caption{Results on the Easy and Hard subsets for
Cantonese (Yue) and Sichuanese (Chuan). Yue uses MER and Chuan uses CER;
both are reported in percent.}
\label{tab:dialect_results}

\small
\setlength{\tabcolsep}{6pt}
\renewcommand{\arraystretch}{1.17}

\begin{adjustbox}{max width=\textwidth}
\begin{tabular}{@{}lllrrrr@{}}
\toprule
Dialect & TTS Model & ASR Model
& \multicolumn{2}{c}{Easy}
& \multicolumn{2}{c}{Hard} \\
\cmidrule(lr){4-5} \cmidrule(l){6-7}
& & & Error$\downarrow$ & SIM$\uparrow$
    & Error$\downarrow$ & SIM$\uparrow$ \\
\midrule

\multirow{3}{*}{Yue}
& M5-TTS (GRPO)
& U2pp-Conformer-Yue
& \textbf{8.55} & 0.47 & \textbf{8.70} & 0.51 \\

& CosyVoice2-Yue
& U2pp-Conformer-Yue
& 10.33 & \textbf{0.62} & 9.49 & \textbf{0.67} \\

& Llasa-1B-Yue
& U2pp-Conformer-Yue
& 10.89 & 0.51 & 12.78 & 0.55 \\

\midrule

\multirow{2}{*}{Chuan}
& M5-TTS (GRPO)
& FireRedASR-AED
& 5.08 & 0.58 & \textbf{7.78} & 0.59 \\

& CosyVoice2-WSC
& FireRedASR-AED
& \textbf{4.28} & \textbf{0.64} & 8.78 & \textbf{0.65} \\

\bottomrule
\end{tabular}
\end{adjustbox}
\end{table*}

\FloatBarrier
\subsection{VoiceJudge-Based Naturalness Evaluation}
\label{sec:voicejudge_eval}

\paragraph{A learned proxy for human assessment.}
VoiceJudge is an internally developed reward model trained on
approximately 100,000 human-annotated speech preference pairs.
Its architecture combines self-supervised learning (SSL) speech
representations with a Transformer-based scoring module.
The model supports multidimensional A/B assessment, including
naturalness and expressiveness, allowing repeated comparisons without
collecting new listener ratings for every experiment.
Internal evaluations indicate good agreement with human judgments.
In this experiment, we use only the naturalness dimension.
The evaluation audio is strictly deduplicated against the preference
training data, with no overlap between the two sets.
We plan to release VoiceJudge model weights and inference code.

\paragraph{Paired evaluation protocol.}
We evaluate 311 pairs of M5-TTS and MOSS-TTS-Nano outputs generated
from the same target text within each pair.
Each pair is assessed in two presentation orders: first with
MOSS-TTS-Nano as A and M5-TTS as B, and then with the order reversed.
A/B decisions are mapped back to the actual system identities before
aggregation.

\begin{table}[!htbp]
\centering
\caption{VoiceJudge-based naturalness preferences on the same 311
audio pairs under two presentation orders.
Reward values are scalar outputs of VoiceJudge.}
\label{tab:voicejudge_naturalness}
\small
\setlength{\tabcolsep}{7pt}
\renewcommand{\arraystretch}{1.15}
\begin{tabular}{@{}lcc@{}}
\toprule
Metric
& \makecell{Original order\\A: MOSS; B: M5}
& \makecell{Swapped order\\A: M5; B: MOSS} \\
\midrule
Evaluated pairs & 311 & 311 \\
M5-TTS wins & \textbf{264 (84.89\%)} & \textbf{269 (86.50\%)} \\
MOSS-TTS-Nano wins & 47 (15.11\%) & 42 (13.50\%) \\
Ties & 0 & 0 \\
Mean reward: M5-TTS & 0.2586 & 0.3231 \\
Mean reward: MOSS-TTS-Nano & $-0.6542$ & $-0.6234$ \\
\bottomrule
\end{tabular}
\end{table}

\paragraph{Naturalness preferences and order consistency.}
As shown in Table~\ref{tab:voicejudge_naturalness}, VoiceJudge prefers
M5-TTS in 264 of 311 pairs (84.89\%) in the original order and
269 pairs (86.50\%) after swapping the order.
The corresponding mean rewards are 0.2586 and 0.3231 for M5-TTS,
versus $-0.6542$ and $-0.6234$ for MOSS-TTS-Nano.
The M5-TTS win rate increases by 1.61 percentage points after swapping.
The preferred system remains unchanged for 288 of 311 pairs (92.60\%):
255 pairs favor M5-TTS in both orders and 33 favor MOSS-TTS-Nano
in both orders (Table~\ref{tab:voicejudge_order}).
The other 23 pairs (7.40\%) change preference.
Thus, the aggregate preference favors M5-TTS under both tested orders,
although some individual decisions change under order reversal.

\begin{table}[!htbp]
\centering
\caption{Joint VoiceJudge preferences after mapping A/B choices to
system identities. The first four rows partition the 311 audio pairs;
the final row summarizes agreement across orders.}
\label{tab:voicejudge_order}
\small
\setlength{\tabcolsep}{8pt}
\renewcommand{\arraystretch}{1.15}
\begin{tabular}{@{}llrr@{}}
\toprule
Original preference & Swapped preference & Pairs & Percentage \\
\midrule
M5-TTS & M5-TTS & 255 & 81.99\% \\
MOSS-TTS-Nano & MOSS-TTS-Nano & 33 & 10.61\% \\
MOSS-TTS-Nano & M5-TTS & 14 & 4.50\% \\
M5-TTS & MOSS-TTS-Nano & 9 & 2.89\% \\
\midrule
\multicolumn{2}{l}{Same preferred system (subtotal)}
& \textbf{288} & \textbf{92.60\%} \\
\bottomrule
\end{tabular}
\end{table}

\FloatBarrier
\subsection{Efficiency Evaluation}
\label{sec:cpu_eval}
We evaluate the inference efficiency of M5-TTS using two key metrics: first-chunk latency and real-time factor (RTF). First-chunk latency measures the elapsed time from the initiation of speech generation to the availability of the first audio chunk, reflecting the system's responsiveness in streaming synthesis. RTF is defined as the ratio of total synthesis time to the duration of the generated speech, measuring the overall computational efficiency of speech generation. An RTF below 1 indicates that speech can be synthesized faster than its playback duration, a key requirement for real-time applications.

On four cores of an Intel Xeon Gold 6530, M5-TTS achieves an RTF of 0.653 and a warm first-chunk latency of 225\,ms. This corresponds to approximately 1.53 seconds of generated speech per second of computation. The timing excludes model loading and warmup and demonstrates faster-than-real-time streaming in the evaluated server-CPU configuration; it does not establish performance on mobile or ARM devices.

\section{Limitations and Future Work}

The current 16-codebook speech representation introduces considerable
inference overhead in the local autoregressive decoder.
Repeated codebook-generation steps require frequent model-weight and cache
accesses and can place substantial demands on memory bandwidth,
limiting CPU inference efficiency despite the compact LM.
We plan to investigate inference with eight codebooks to reduce local
decoding overhead, while evaluating the resulting trade-off between
computational efficiency and acoustic fidelity.

Audio quality also remains a limitation.
The current system supports only 24\,kHz output, and highly expressive
synthesis can exhibit voice breaks and hoarse or muted segments.
These artifacts reduce perceived audio quality and indicate that
expressive acoustic generation requires further refinement.
Future work will extend the system to support 48\,kHz output and
improve the robustness of acoustic generation under expressive speaking
conditions.

Intelligibility remains challenging on difficult inputs.
M5-TTS (GRPO) achieves a CER of 17.48\% on Seed-Hard, compared with
15.77\% for MOSS-TTS-Nano, indicating that its pronunciation robustness
under challenging conditions is still insufficient.
Addressing this gap requires continued refinement of the data pipeline
and training-data mixture.
We will improve data quality control, expand coverage of challenging
pronunciation patterns, and refine sampling proportions across general
and difficult examples to strengthen robustness while preserving
performance on standard multilingual inputs.

We also plan to introduce explicit control over emotion,
paralinguistic expressions, and speaking style, including controllable
emotional delivery and nonverbal vocalizations such as laughter and sighs.
These extensions aim to enable more expressive and context-appropriate
speech generation for on-device applications and embodied voice interaction.

\section{Conclusion}

We presented M5-TTS, a compact speech generation framework built around
a 50M-parameter language model, excluding the speech codec.
Its hierarchical architecture combines a 16-codebook speech representation
with multi-token prediction. The system supports multilingual and
multi-dialect synthesis, reference-based zero-shot voice cloning, and
speaker-specific multilingual generation.
GRPO post-training combines separately group-normalized speaker-similarity
and pause advantages with CER-guided local process feedback.
Character edit counts yield group-relative advantages; negative feedback
propagates backward within a clause before projection onto codec frames.
The resulting local process advantage is added to the global advantage
and shared by all codec tokens in a frame for policy optimization.

Experiments demonstrate strong content accuracy within this compact
parameter budget. Using half the LM parameters of MOSS-TTS-Nano,
M5-TTS achieves relative error reductions of 62.2\% on Seed-ZH and
35.3\% on Seed-EN, and lower content error rates in five of the six
MiniMax languages.
Compared with CPT, GRPO reduces content error rates across the reported
Seed and MiniMax subsets while maintaining or improving speaker
similarity at the reported precision.
Dialect evaluations further show lower Cantonese MER than both compared
baselines and improved performance on the Sichuanese Hard subset,
although challenging Chinese utterances and dialectal speaker
preservation remain areas for improvement.

We also introduced VoiceJudge, an SSL--Transformer reward model trained
on approximately 100,000 human-annotated preference pairs, for scalable
model-based perceptual assessment. On 311 audio pairs disjoint from
VoiceJudge's preference-training data, its naturalness dimension prefers
M5-TTS over MOSS-TTS-Nano in
84.89\% and 86.50\% of comparisons under two A/B orders, with 92.60\%
agreement on the winning system across orders.
We plan to release VoiceJudge model weights and inference code for further
validation and community use.

On four cores of an Intel Xeon Gold 6530, M5-TTS achieves an RTF of
0.653 and a warm first-chunk latency of 225\,ms, demonstrating
faster-than-real-time synthesis with responsive streaming output
in the evaluated CPU configuration.
These results support M5-TTS as a compact speech generation component
for interactive applications and motivate further deployment and
evaluation on resource-constrained edge devices for embodied
voice interaction.

\section{Ethics Statement}

Voice cloning and speaker-specific synthesis can support accessibility, language learning, and personalized interfaces, but can also facilitate impersonation, fraud, harassment, or misleading recordings. Use of M5-TTS should require appropriate authorization from the represented speaker and respect applicable identity, publicity, and intellectual-property rights. Synthetic speech should be disclosed where listeners could otherwise reasonably mistake it for an authentic recording. The system should not be used to attribute statements to a person deceptively or to evade identity verification.

Training data and reference recordings raise separate licensing and privacy concerns. Data use and redistribution should follow the permissions and restrictions of each source, including those governing derived checkpoints. Reference audio, transcripts, and speaker representations should be handled as potentially identifying information, with collection, retention, and sharing limited to their stated purpose. Local inference can reduce transmission to external services, but does not by itself guarantee privacy.

Coverage and synthesis quality may differ across languages, dialects, accents, and speaker groups. The reported benchmark results should not be treated as proof of equal performance or suitability for consequential applications. 

\clearpage
\appendix
\section{Implementation and Reproducibility Details}
\label{sec:appendix}

\subsection{Conditioning and Codebook Dependencies}
\label{app:conditioning}

In the speaker-specific setting, the conditioning prefix orders accent,
speaker, and language identifiers before the text. Pre-training omits the
speaker field. The following listing preserves the special-token spellings;
line breaks are removed when concatenating the prefix. Braced fields denote
values to substitute, rather than literal text to synthesize.

\begin{lstlisting}[basicstyle=\ttfamily\footnotesize,columns=fullflexible,
  keepspaces=true,showstringspaces=false,breaklines=true,frame=single]
<|accent_begin|>{accent id}<|accent_end|>
<|speaker_start|>{speaker id}<|speaker_end|>
<|lang_begin|>{lang id}<|lang_end|>{text}
\end{lstlisting}

This prefix specifies the ordering of the three controls; it does not specify
the full serialization of reference speech for voice cloning. Identifier
inventories, text normalization, and reference-prompt construction should
accompany the released inference implementation. No additional codec boundary
tokens are implied by this notation.

Let $q_{t,c}$ denote codebook $c$ at audio frame $t$, with $c\in\{0,\ldots,15\}$.
The temporal backbone consumes the conditioning sequence and completed earlier
frames. Within a historical frame, the 16 codebook embeddings are summed into
one frame-level input. Its hidden state $h_t$ first predicts $q_{t,0}$; the local
MTP Transformer subsequently generates $q_{t,1},\ldots,q_{t,15}$ sequentially:
\begin{equation}
 p(\mathbf{q}_t\mid h_t)
 =p(q_{t,0}\mid h_t)
  \prod_{c=1}^{15}p(q_{t,c}\mid h_t,q_{t,0:c-1}).
\end{equation}
Thus, the residual codebooks are not 15 independent parallel predictions.
The backbone cache persists across frames, whereas the local cache is reset
for each new frame. The local Transformer predicts codec indices; the separate
speech-codec decoder reconstructs the waveform. The 50M parameter count refers
to the LM component and excludes that codec.

\subsection{Pre-training and Continued Pre-training Hyperparameters}
\label{app:training_hyperparameters}

Table~\ref{tab:pretrain_cpt_hyperparameters} summarizes the configurations
used for pre-training and continued pre-training (CPT).
Both stages share the same backbone, local decoder, tokenizer, and
16-codebook speech representation, and use BF16 precision on CUDA.
CPT uses lower learning rates, a shorter learning-rate schedule,
and a larger maximum token budget. We use the Llama 3 text
tokenizer~\cite{llama3_2024}, recorded as \texttt{llama3} in the
configuration. The speech tokenizer is described in
Section~\ref{sec:architecture}.

\begin{table}[!htbp]
\centering
\caption{Hyperparameter configurations for pre-training and CPT.
Field names follow the training configuration.}
\label{tab:pretrain_cpt_hyperparameters}

\small
\setlength{\tabcolsep}{10pt}
\renewcommand{\arraystretch}{1.17}

\begin{tabular}{@{}lcc@{}}
\toprule
Hyperparameter & Pre-training & CPT \\
\midrule

\texttt{backbone\_flavor}
& \texttt{llama-30M} & \texttt{llama-30M} \\

\texttt{decoder\_flavor}
& \texttt{llama-20M} & \texttt{llama-20M} \\

\texttt{tokenizer}
& \texttt{llama3} & \texttt{llama3} \\

\texttt{audio\_num\_codebooks}
& 16 & 16 \\

\midrule

\texttt{learning\_rate}
& $1 \times 10^{-4}$ & $1 \times 10^{-5}$ \\

\texttt{min\_lr}
& $1 \times 10^{-5}$ & $1 \times 10^{-6}$ \\

\texttt{scheduler\_steps}
& 1,000,000 & 150,000 \\

\texttt{num\_warmup\_steps}
& 4,000 & 1,000 \\

\texttt{save\_every}
& 500,000 & 15,000 \\

\midrule

\texttt{max\_tokens}
& 13,000 & 25,000 \\

\texttt{max\_batch\_size}
& 999 & 999 \\

\texttt{pad\_multiple}
& 16 & 16 \\

\texttt{decoder\_fraction}
& 0.125 & 0.125 \\

\texttt{lambda\_depth}
& 1.0 & 1.0 \\

\midrule

\texttt{device}
& \texttt{cuda} & \texttt{cuda} \\

\texttt{precision}
& \texttt{bf16} & \texttt{bf16} \\

\bottomrule
\end{tabular}
\end{table}

\FloatBarrier
\subsection{GRPO Reward and Loss Details}
\label{app:grpo_details}

\paragraph{Pause score.}
For adjacent aligned characters, let $s_{i,j}$ and $u_{i,j}$ denote the
start and end timestamps of character $j$ in candidate $i$. The gap is
$d_{i,j}=\max(0,s_{i,j+1}-u_{i,j})$. Ordinary boundaries and boundaries
associated with punctuation use different duration constraints:
\begin{equation}
    r_{i,j}^{\mathrm{pause}}
    =\begin{cases}
      \mathbb{I}(d_{i,j}\leq\tau_{\mathrm{intra}}),
         &j\in\mathcal{B}_{i,\mathrm{intra}},\\
      \mathbb{I}(\tau_{\mathrm{punct}}^{-}\leq d_{i,j}
                       \leq\tau_{\mathrm{punct}}^{+}),
         &j\in\mathcal{B}_{i,\mathrm{punct}}.
    \end{cases}
\label{eq:pause-boundary-score}
\end{equation}
The utterance score averages these indicators over the valid aligned
boundaries $\mathcal{B}_i$:
\begin{equation}
    R_i^{\mathrm{pause}}
       =\frac{1}{|\mathcal{B}_i|}
          \sum_{j\in\mathcal{B}_i}r_{i,j}^{\mathrm{pause}}.
\label{eq:pause-utterance-score}
\end{equation}
The thresholds define the allowed intra-sentence gap and the acceptable
punctuation-pause range. This score is an utterance-level reward; local
content feedback is constructed separately from character edit counts.

\paragraph{Illustrative credit propagation.}
Consider four characters in one clause with unpropagated advantages
$(a_{i,1},a_{i,2},a_{i,3},a_{i,4})=(0,0,-1,0)$.
Equation~\ref{eq:grpo-char-propagation} gives
$\widetilde{\mathbf{a}}_i=(-e^{-2},-e^{-1},-1,0)$.
For frames fully contained in the respective character intervals, the
local contributions added to the global advantage are approximately
$(-0.068,-0.184,-0.5,0)$ at $\lambda_{\mathrm{CER}}=0.5$.
The first two characters receive attenuated feedback from the third,
although their original local advantages were zero. The fourth character
receives no local contribution in this example. This illustrates credit
propagation; the values are not measured training results.

\paragraph{Multi-codebook loss aggregation.}
The first-codebook policy loss is normalized by the number of generated
decisions. Local-decoder losses are weighted across codebooks on sampled
frames and normalized by the sampled-frame count. The two components are
added, and the same aggregation is applied to the token-wise KL estimates.
Valid reference-prefix positions have zero policy advantage and contribute
only to KL regularization. No length-based scaling or additional clipping
is applied to the local process advantage, as recorded in
Table~\ref{tab:grpo_hyperparameters}.

\subsection{GRPO Hyperparameters}
\label{app:grpo_hyperparameters}

Table~\ref{tab:grpo_hyperparameters} summarizes the GRPO configuration.
Each global batch contains 32 prompts, with eight candidates generated
per prompt, yielding 256 candidate utterances.
All eight candidates are retained without rollout selection.
Utterance-level SIM and pause rewards are normalized separately within
each prompt group, then weighted by 1.0 and 0.5, respectively, and summed
without further normalization. Utterance-level CER broadcasting is disabled;
the projected local process advantage is added separately with a
weight of 0.5, following Equation~\ref{eq:grpo-frame-advantage}. The KL
regularization coefficient is 0.01. The \texttt{token\_cer.*} fields retain
their implementation names and configure this CER-guided local process
feedback: errors are compared at text-character positions and projected
onto codec frames, whose 16 codec tokens share an advantage.

\begin{table}[tbp]
\centering
\caption{Hyperparameters for GRPO post-training.
Field names follow the training configuration.}
\label{tab:grpo_hyperparameters}

\small
\setlength{\tabcolsep}{10pt}
\renewcommand{\arraystretch}{1.08}

\begin{tabular}{@{}lc@{}}
\toprule
Hyperparameter & Value \\
\midrule
\multicolumn{2}{@{}l}{\textit{Rewards and optimization}} \\
\texttt{reward\_weight.cer\_reward}   & 0.0 \\
\texttt{reward\_weight.sim\_reward}   & 1.0 \\
\texttt{reward\_weight.pause\_reward} & 0.5 \\
\texttt{token\_cer.enabled}          & \texttt{true} \\
\texttt{token\_cer.mode}             & \texttt{char\_exp} \\
\texttt{token\_cer.weight}           & 0.5 \\
\texttt{token\_cer.length\_scale}     & \texttt{false} \\
\texttt{token\_cer.advantage\_clip}   & \texttt{null} \\
\texttt{token\_cer.ambiguous}        & \texttt{use\_path} \\
\texttt{use\_gdpo}                  & \texttt{false} \\
\texttt{kl\_weight}                 & 0.01 \\

\midrule
\multicolumn{2}{@{}l}{\textit{Batching and rollout}} \\
\texttt{batch\_size}                    & 32 \\
\texttt{batch\_size\_scope}              & \texttt{global} \\
\texttt{batch\_size\_per\_rank}          & 16 \\
\texttt{num\_answers\_per\_question}     & 8 \\
\texttt{rollout\_micro\_batch\_size}     & 8 \\
\texttt{rollout\_selection.enabled}     & \texttt{false} \\
\texttt{rollout\_selection.keep\_per\_group} & 8 \\
\texttt{rollout\_mode}                  & \texttt{remote} \\
\texttt{async\_rollout}                 & \texttt{true} \\
\texttt{max\_policy\_lag}               & 1 \\
\texttt{resample\_max\_attempts}        & 3 \\

\midrule
\multicolumn{2}{@{}l}{\textit{Generation settings}} \\
\texttt{generation\_conf.temperature}          & 1.0 \\
\texttt{generation\_conf.topk}                 & 50 \\
\texttt{generation\_conf.max\_audio\_length\_ms} & 40,000 \\
\texttt{generation\_conf.max\_seq\_len}        & 2,048 \\

\midrule
\multicolumn{2}{@{}l}{\textit{Reward execution and reproducibility}} \\
\texttt{async\_reward}          & \texttt{true} \\
\texttt{reward\_workers}        & 8 \\
\texttt{reward\_max\_inflight}   & 16 \\
\texttt{reward\_timeout}        & 300 \\
\texttt{seed}                  & 10,086 \\
\bottomrule
\end{tabular}

\par\vspace{3pt}
\begin{minipage}{\linewidth}
\footnotesize
\textit{Notes.} Reward workers are HTTP threads per actor.
Rollout selection is disabled, so its configured selection method is inactive.
\end{minipage}
\end{table}

\clearpage
\bibliography{biblio}
\bibliographystyle{colm2024_conference}

\end{document}